\documentclass[12pt]{article}
\usepackage{amssymb}
\makeatletter \@addtoreset{equation}{section}

\makeatother
\newcommand{\fft}[2]{{\frac{#1}{#2}}}
\newcommand{\ft}[2]{{\textstyle\frac{#1}{#2}}}
\newcommand{\R}{{\mathbb R}}

\newcommand{\Hol}{{\mathrm{Hol}}}
\newcommand{\SO}{{\mathrm{SO}}}
\newcommand{\SL}{{\mathrm{SL}}}
\newcommand{\SU}{{\mathrm{SU}}}
\newcommand{\D}{{\mathcal D}}
\newcommand{\Rm}{{\mathcal R}}
\newcommand{\Cm}{{\mathcal C}}
\begin{document}

\thispagestyle{empty}
\addtocounter{page}{-1}

\begin{flushright}
MCTP-04-69\\[-6pt]
FTUV-04-1214\\[-6pt]
IFIC-04-71\\[-6pt]
hep-th/0412154
\end{flushright}

\vspace*{1cm}

\begin{center}

{\large\bf Higher Order Integrability in Generalized Holonomy}
\vspace*{.5cm}

A. Batrachenko$^\dagger$, James T. Liu$^\dagger$, Oscar Varela$^{\dagger,*}$
and W. Y. Wen$^\dagger$\footnote{E-mails: abat@umich.edu, jimliu@umich.edu,
oscar.varela@ific.uv.es, wenw@umich.edu}

\vspace*{1cm}
{\it {}$^\dagger$Michigan Center for Theoretical Physics,\\
Randall Laboratory, Department of Physics, University of Michigan,\\
Ann Arbor, MI 48109--1120, USA\\}
\vspace*{.5cm}
{\it {}$^*$Departamento de F\'\i{}sica Te\'orica, Universidad de Valencia
and IFIC (CSIC-UVEG),\\
46100-Burjassot (Valencia), Spain\\}
\vspace*{1cm}

\underline{ABSTRACT}

\end{center}

Supersymmetric backgrounds in M-theory often involve four-form flux in
addition to pure geometry.  In such cases, the classification of
supersymmetric vacua involves the notion of generalized holonomy
taking values in $\SL(32,\R)$, the Clifford group for eleven-dimensional
spinors.  Although previous investigations of generalized holonomy
have focused on the curvature $\Rm_{MN}(\Omega)$ of the generalized
$\SL(32,\R)$ connection $\Omega_M$, we demonstrate that this local
information is incomplete, and that satisfying the higher order
integrability conditions is an essential feature of generalized holonomy.
We also show that, while this result differs from the case of ordinary
Riemannian holonomy, it is nevertheless compatible with the Ambrose-Singer
holonomy theorem.

\newpage

\section{Introduction}

The connection between holonomy and supersymmetry is a close and important
one.  The notion of a supersymmetric background is simply one where some
fraction of the supersymmetry variations vanish.  In particular, for bosonic
field configurations, we are invariably led to the vanishing of the
gravitino transformation
\begin{equation}
\delta\psi_M\equiv\D_M\epsilon=0,
\label{eq:kse}
\end{equation}
where $\D_M$ is the supercovariant derivative.  The number of
preserved supersymmetries is then equal to the number of linearly
independent solutions of the Killing spinor equation (\ref{eq:kse}).
Thus the goal of enumerating supersymmetric vacua is essentially one
of classifying all solutions to the above Killing spinor equation.

A necessary condition for the existence of Killing spinors is obtained
from the integrability of the Killing spinor equation (\ref{eq:kse})
\begin{equation}
M_{MN}\epsilon\equiv[\D_M,\D_N]\epsilon=0.
\label{eq:intc}
\end{equation}
However, it ought to be evident that this simply measures the effect
of parallel transportation of a spinor around an infinitesimal loop
along the $M$ and $N$ directions of the base manifold.  By the Ambrose-Singer
theorem \cite{ambrose}, this is in general related to the Lie algebra of
some holonomy group $\Hol(\D)$ acting on the
spinors.  For the case of a Riemannian connection, so that $\D_M$ is
identified with $\nabla_M$, the gravitational covariant derivative,
the first order integrability condition directly yields the conventional
Riemannian
holonomy group $\Hol(\nabla)\subseteq \SO(n)$ where $\SO(n)$
is the Riemannian structure group for an $n$-dimensional orientable
Euclidean manifold with a metric.  In this case, the analysis is quite
familiar, and holonomy groups have been classified by Berger in
\cite{Berger} for the Euclidean case, and partially extended to the
Lorentzian case by Bryant in \cite{Bryant}.

In practice, in order to obtain Killing spinors, one often starts
with the integrability condition (\ref{eq:intc}) and not directly with
the Killing spinor equation (\ref{eq:kse}), as the integrability condition
is only algebraic in $\epsilon$.  For the case of a Riemannian
connection, use of the Ambrose-Singer theorem demonstrates that the
integrability condition (\ref{eq:intc}) is also a sufficient condition for
the existence of Killing spinors.  However, it has been observed that
this is no longer the case for more general connections.  This is perhaps
most evident in the squashed 7-sphere compactification of M-theory
\cite{ADP,DNP83}, where left-squashing preserves $N=1$ supersymmetry in four
dimensions while right-squashing leaves no unbroken supersymmetries at all.
Yet, at the same time, first order integrability of the form (\ref{eq:intc})
cannot distinguish between the two cases; only by going to second order
integrability can the issue of the existence or non-existence of Killing
spinors be resolved \cite{vNW}.

This question of whether the algebra generated by the curvature [expressed
in the first order integrability condition (\ref{eq:intc})] agrees or does
not agree with the algebra of the holonomy group has until now been mostly
ignored in the study of generalized holonomy
\cite{Duff:2003ec,Hull:2003mf,Papadopoulos:2003pf}.  At present we will
focus on generalized holonomy in eleven-dimensional supergravity.  In this
case, the bosonic fields are the metric $g_{MN}$ and $3$-form potential
with $4$-form field strength $F_{(4)}$.  The fermionic superpartner is
simply the gravitino, with transformation
\begin{equation}
\D_M\equiv\partial_M+\ft14\Omega_M=
\nabla_M-\ft1{288}(\Gamma_M{}^{NPQR}-8\delta_M^N\Gamma^{PQR})
F_{NPQR}.
\label{eq:gencd}
\end{equation}
Here $\Omega_M$ is considered to be a generalized connection, consisting
of the conventional Riemannian connection as well as the flux-induced
term and taking values in the space of forms $\Lambda^*(\R^{1,10})$
(which is identical to the Clifford algebra of the Dirac matrices).
Actually it is only the even part of the Clifford group that is relevant;
as a result the generalized structure group is $\mathrm{SL}(32,\R)$
\cite{Hull:2003mf}, which is considerably larger than the Riemannian
structure group $\SO(1,10)$.

The idea behind generalized holonomy is simply to consider the holonomy
of the generalized connection $\Omega_M$.  Since $\Omega_M$ takes
values in the generalized structure group, we see that the generalized
holonomy group is a subgroup of $\mathrm{SL}(32,\R)$.  Furthermore,
as shown in \cite{Hull:2003mf}, for a background preserving $n$
supersymmetries, the generalized holonomy must be contained in
$\mathrm{SL}(32-n,\R)\ltimes n\R^{32-n}$.  As a result, the
issue of classifying supersymmetric vacua may be mapped into one of
classifying the generalized holonomy groups as subgroups of $\mathrm{SL}
(32,\R)$. Expressions for the generalized curvature of a background
preserving $n$
supersymmetries were given in \cite{BAIPV} (including the conjectured
preonic case \cite{BAIL}, with $n=31$) by relating the
notions of Killing and preonic spinors, and
an investigation of basic supersymmetric configurations
of M-theory was performed in \cite{BDLW}, where a large variety of
generalized holonomy groups were obtained.  However, one of the striking
results of the analysis of \cite{BDLW} was the fact that identical
generalized holonomies may yield different amounts of supersymmetries.
This shows that knowledge of the holonomy group is insufficient to fully
classify the background, and that knowledge of the decomposition of the
$32$-component spinor under $\Hol(\D)$ is also needed.

At a more technical level, it was also seen in \cite{BDLW} that in many
cases the complete Lie algebra of $\Hol(\D)$ was not obtained
from first order integrability (\ref{eq:intc}), so that in particular the
algebra had to be closed by hand.  This issue is rather suggestive that
the generalized curvature at a local point carries incomplete information
of the generalized holonomy group, in apparent violation of the
Ambrose-Singer theorem (but in agreement with the issue of left- versus
right-squashing of $S^7$ mentioned above).  However a careful reading
of the Ambrose-Singer theorem indicates that $\Hol_p(\D)$
at a point $p$ is spanned by elements of the generalized curvature
(\ref{eq:intc}) not just at point $p$, but at all points $q$ connected
to $p$ by parallel transport (see {\it e.g.} \cite{nakahara},~p.~388 or
\cite{joyce},~p.~33).  Thus there is in fact no contradiction.  Furthermore,
this is rather suggestive that satisfying higher order integrability
(representing motion from $p$ to $q$) is in fact a necessary condition for
identifying the proper generalized holonomy group, and that it is the
Riemannian case that is the exception.

These issues have led us to the present work, where we explore the interplay
of higher order integrability and generalized holonomy.  We begin by
revisiting the generalized holonomy of the M5-brane and M2-brane solutions
of supergravity, and show that higher order integrability yields precisely
the `missing' generators that were needed to close the algebra.  Other
than this, however, the generalized holonomy groups $\SO(5)\ltimes
6\R^{4(4)}$ for the M5-brane and $\SO(8)\ltimes12\R^{2(8_s)}$ for the
M2-brane identified in \cite{BDLW} are unchanged.  Following this, we
turn to the squashed $S^7$ \cite{ADP,DNP83}, where the situation is
considerably different.

The importance of
higher order integrability was of course previously recognized in \cite{vNW}
for the case of the squashed $S^7$.  Here, we reinterpret the result of
\cite{vNW} in the language of generalized holonomy, and confirm the
statement of \cite{Duff:2002rw}
that while first order integrability yields the incorrect result
$\Hol^{(1)}(\D)=G_2\subset\SO(7)\subset\SO(8)$,
higher order integrability corrects this to $\Hol(\D)=
\SO_\pm(7)\subset\SO(8)$, where the two distinct possibilities
$\SO_-(7)$ and $\SO_+(7)$ arise from left- and right-squashing,
respectively.  Since the spinor decomposes as either $\mathbf 8_s \to
\mathbf 7 + \mathbf 1$ or $\mathbf 8_s \to \mathbf 8$ in
the two cases, this explains the resulting $N=1$ or $N=0$ supersymmetry in
four dimensions \cite{Duff:2002rw}.

In the following section, we provide a brief review of higher order
integrability and then proceed to reexamine the generalized holonomy
of the M5-brane and M2-brane solutions of supergravity.  For both
cases, we find that the higher order conditions close the holonomy
algebra but otherwise do not affect the results of \cite{BDLW}.
In section~3 we turn to the squashed $S^7$ example and show that for
this case higher order integrability is essential in obtaining the
correct holonomy group from the curvature.  Some of the details of
this example are relegated to an Appendix.

\section{Generalized curvature and higher order integrability}

We start by defining the generalized curvature and $n$-th order
integrability conditions.  For a generalized covariant derivative of
the form (\ref{eq:gencd})
\begin{equation}
\D_M\equiv\partial_M+\ft14\Omega_M,
\label{eq:omega}
\end{equation}
first order integrability (\ref{eq:intc}) yields
\begin{equation}
M_{MN}\equiv [\D_M(\Omega),\D_N(\Omega)]
=\ft14(\partial_M\Omega_N-\partial_N\Omega_M
+\ft14[\Omega_M,\Omega_N])\equiv\ft14\Rm_{MN}(\Omega),
\label{eq:1int}
\end{equation}
where $\Rm_{MN}(\Omega)$ is the curvature of the generalized connection
$\Omega$; in particular, $\Rm_{MN}(\Omega)=R_{MNPQ}\Gamma^{PQ}+\cdots$.
It is a familiar result that, when contracted with $\Gamma^M$, the first
order integrability condition $\Gamma^M\Rm_{MN}(\Omega)\epsilon=0$ yields
an expression compatible with the bosonic equations of motion
\cite{GP,BAIPV}, which for eleven-dimensional supergravity read
\begin{eqnarray}
\label{Einstein}
&& R_{MN}=\ft1{12}\left(F_{MPQR}F_N{}^{PQR}-\ft1{12}g_{MN}
F^{PQRS}F_{PQRS}\right),\\
&& d*F_{(4)}+\ft12 F_{(4)} \wedge F_{(4)}=0.
\label{4form}
\end{eqnarray}

Higher order integrability expressions may be obtained by taking generalized
covariant derivatives of (\ref{eq:1int}).  Here we make the definition
precise by taking the chain of expressions
\begin{eqnarray}
M_{AMN} &\equiv& [\D_A,M_{MN}] = \ft14\D_A(\Omega)\Rm_{MN}(\Omega),
\label{eq:2int}\\
M_{ABMN} &\equiv& [\D_A,M_{BMN}]=\ft14\D_A(\Omega)\D_B(\Omega)\Rm_{MN}(\Omega).
\label{eq:3int}\\
&\vdots&\nonumber
\end{eqnarray}
It ought to be evident that the higher order integrability conditions
correspond to measuring the generalized curvature $\Rm_{MN}(\Omega)$
parallel transported away from the original base point $p$.  In this
sense, the information obtained from higher order integrability is
precisely that required by the Ambrose-Singer theorem in making the
connection between $\Hol_p(\D)$ and the curvature of the generalized
connection.

Note that we take $\Omega_M$ in (\ref{eq:omega}) to include the
Levi-Civita connection on the base in addition to the generalized
connection on the Clifford bundle.  In this sense, we actually work
with a generalized connection in $\mathcal TM\times
\mathrm{Cliff}(M)$.  Writing
\begin{equation}
\D_M=\partial_M+\ft14\Omega_M=\nabla_M+\ft14\widetilde\Omega_M,
\end{equation}
where
\begin{equation}
\widetilde\Omega_M=-\ft1{72}(\Gamma_M{}^{NPQR}-8\delta_M^N\Gamma^{PQR})
F_{NPQR}
\end{equation}
in the case of eleven-dimensional supergravity, the integrability
expressions (\ref{eq:1int}), (\ref{eq:2int}) and (\ref{eq:3int})
are equivalent to
\begin{eqnarray}
M_{MN}&=&\ft14\Rm_{MN}(\Omega)=\ft14R_{MNAB}\Gamma^{AB}+\ft12\nabla_{[M}
\widetilde\Omega_{N]}+\ft19\widetilde\Omega_{[M}\widetilde\Omega_{N]},
\nonumber\\
M_{AMN}&=&\D_A(\Omega)M_{MN}=\nabla_AM_{MN}+\ft14[\widetilde\Omega_A,M_{MN}],
\nonumber\\
M_{ABMN}&=&\D_A(\Omega)M_{BMN}=\nabla_AM_{BMN}
+\ft14[\widetilde\Omega_A,M_{BMN}],\nonumber\\
&\vdots&
\label{eq:nhint}
\end{eqnarray}
We occasionally find these expressions useful for direct computation.

\subsection{Generalized holonomy of the M5-brane}

As examples of how higher order integrability may affect determination
of the generalized holonomy group, we first revisit the case of the M5-
and M2-brane solutions of supergravity.  The generalized holonomy of
these solutions, as well as several others, was originally investigated
in \cite{BDLW}.  For vacua with non-vanishing flux, including
the brane solutions, it was seen that the Lie algebra generators obtained
from first order integrability, (\ref{eq:1int}), are insufficient for
closure of the algebra.  In particular, additional generators must be
obtained by further commutators.  In \cite{BDLW}, this was done by
closing the algebra by hand.  In the present context, however, additional
commutators are readily available from the higher order integrability
expressions, (\ref{eq:nhint}).

It turns out that, for the M5-brane, working up the third order integrability
is sufficient to close the algebra.  To see this, we recall the M5-brane
solution is given in isotropic coordinates as
\begin{eqnarray}
&&ds^2_{11}=H_5^{-1/3}dx_{\mu}^2+H_5^{2/3}d\vec{y}\,^2,\nonumber\\
&&F_{ijkl}=\epsilon_{ijklm}\partial_m H_5,
\end{eqnarray}
where $H_5(\vec{y}\,)$ is harmonic in the six-dimensional transverse
space spanned by the $\{y^i\}$, and $\epsilon_{ijklm}=\pm1$ is a density.
Computation of the supercovariant derivative (\ref{eq:gencd}) in this
background yields a generalized connection of the form \cite{BDLW}
\begin{equation}
\Omega_\mu=\Omega_\mu^{\nu i}K_{\mu i},\qquad
\Omega_i=-\ft13\partial_i\ln H_5\Gamma^{(5)}+\ft12\Omega_i^{jk}\hat T_{jk},
\end{equation}
when acting on spinors.  Here we have highlighted the Lie algebra
structure by introducing a set of generators
\begin{equation}
\hat T_{ij}=\Gamma_{\bar i\bar j}P_5^+,\qquad
K_\mu=\Gamma_\mu P_5^+,\qquad
K_{\mu i}=\Gamma_{\bar\mu \bar i}P_5^+,\qquad
K_{\mu ij}=\Gamma_{\bar\mu\bar i\bar j}P_5^+,
\label{eq:m5gen}
\end{equation}
where
$\Gamma^{(5)}\equiv \frac{1}{5!}\epsilon_{ijklm}\Gamma^{\bar i\bar j
\bar k\bar l\bar m}$ and $P_5^+ \equiv \frac{1}{2}(1+\Gamma^{(5)})$
is the half-BPS projector, and the overlined indices refer to local
frame indices.  The component expressions for $\Omega_\mu^{\nu i}$
and $\Omega_i^{jk}$ are
\begin{equation}
\Omega_\mu^{\nu i}= -\ft23H_5^{-1/2}\partial_i \ln H_5,\qquad
\Omega_i^{jk}=\ft83\delta_{i[j}\partial_{k]}\ln{H_5}.
\end{equation}
Even before addressing integrability, we see that $\Omega_M$ includes
the generator $\Gamma^{(5)}$ in addition to $\hat T_{ij}$ and $K_{\mu i}$.
However, the connection itself is not physical, and we see below that
the terms proportional to $\Gamma^{(5)}$ drop out in generalized curvatures
(and hence do not contribute to generalized holonomy).

The first order integrability of the generalized connection, given by
(\ref{eq:1int}), was computed in \cite{BDLW}.  The result was
\begin{eqnarray}
M_{\mu\nu}\equiv\ft14\Rm_{\mu\nu}&=&0,\nonumber\\
M_{\mu i}\equiv\ft14\Rm_{\mu i}&=&H_5^{-1/2}\left[\ft16
(\partial_i\partial_j\ln{H_5}-\ft23\partial_i\ln{H_5}\partial_j\ln{H_5})
+\ft1{18}\delta_{ij}(\partial\ln{H_5})^2\right]K_{\mu j}, \nonumber\\
M_{ij}\equiv\ft14\Rm_{ij}&=&\left[\ft23(\partial_l\partial_{[i\vphantom{]}}
\ln{H_5}
-\ft23\partial_l\ln{H_5}\partial_{[i}\ln{H_5})\delta_{j]k}
-\ft29(\partial\ln H_5)^2\delta_{[i}^k\delta_{j]}^l\right]\hat T_{kl}.
\label{eq:m51order}
\end{eqnarray}
At this point, it was noted that the generators $\hat T_{ij}$ and
$K_{\mu i}$ do not form a closed algebra, as both $K_\mu$ and $K_{\mu ij}$
are missing.  Thus the algebra of the holonomy group, with generators
(\ref{eq:m5gen}), was obtained only after closing the algebra by hand.
In fact we now see that the higher order integrability relations,
expressed as (\ref{eq:nhint}), give rise to a sequence of additional
commutators which are precisely the ones necessary to ensure closure
of the algebra.

For the M5-brane, the second order integrability conditions, defined by
(\ref{eq:2int}), take on the form
\begin{eqnarray}
&&M_{\mu\nu\lambda}=M_{\mu\nu\lambda}^{\rho i}K_{\rho i},\qquad
M_{\mu\nu i}=\ft12M_{\mu\nu i}^{jk}\hat T_{jk},\qquad
M_{\mu ij}=M_{\mu ij}^{\nu k}K_{\nu k}+\ft12M_{\mu ij}^{\nu kl}K_{\nu kl},
\nonumber\\
&&M_{i\mu\nu}=0,\qquad
M_{i\mu j}=M_{i\mu j}^{\nu k}K_{\nu k}+\ft12M_{i\mu j}^{\nu kl}K_{\nu kl},
\qquad
M_{ijk}=\ft12M_{ijk}^{lm}\hat T_{lm},
\label{eq:m5ms}
\end{eqnarray}
where the component factors $M_{AMN}^{\cdots}$ are functions of $H_5$ and its
derivatives.  For example,
\begin{eqnarray}
M_{\mu\nu\lambda}^{\rho i}&=&\ft1{36}H^{-3/2}[\partial_j\ln H_5
\partial_j\partial_i\ln H_5-\ft13\partial_i\ln H_5(\partial H_5)^2
]\eta_{\mu[\nu}\delta_{\lambda]}^\rho,\nonumber\\
M^{jk}_{\mu\nu i} &=&\ft49H^{-1}[\partial_{[j}\ln H_5
\partial^i\partial_{k]}\ln H_5
-\delta_{i[j}\partial^l\ln H_5\partial^l\partial_{k]}\ln H_5]\eta_{\mu\nu}.
\end{eqnarray}
The other factors arising in (\ref{eq:m5ms}) are similar.  However, we
will not need their explicit forms.  We simply note that one additional
generator $K_{\mu ij}$ arises at second order.  However, the algebra is
not closed until third order integrability (\ref{eq:3int}) is taken into
account, since one additional commutator is necessary to provide the
$K_\mu$ generator.  In the case of the M5-brane, no additional information
is gained beyond the third order integrability level.  In fact, the
identification of the proper generalized holonomy group
\begin{equation}
\Hol_{M5}=SO(5)_+ \ltimes 6\R^{4(4)},
\end{equation}
is unchanged from the presentation of \cite{BDLW}.  All that has arisen
from higher order integrability of the M5-brane is closure of the
algebra on the same set of generators that were present at first order.

\subsection{Generalized holonomy of the M2-brane}

The analysis of the M2-brane is similar to that of the M5-brane.  The
supergravity solution takes the form
\begin{eqnarray}
&&ds^2=H_2^{-2/3}dx_\mu^2+H_2^{1/3}d\vec{y}\,^2, \nonumber\\
&&F_{\mu\nu\rho i}=\epsilon_{\mu\nu\rho}\partial_i H_2^{-1},
\end{eqnarray}
where $\epsilon_{\mu\nu\rho}=\pm1$ and $H_2(\vec y\,)$ is a harmonic
function in the transverse space.  Following \cite{BDLW}, we introduce a
set of generators
\begin{equation}
\hat T_{ij}=\Gamma_{\bar i\bar j}P_2^+,\qquad
K_{\mu i}=\Gamma_{\bar\mu\bar i}P_2^+,\qquad
K_{\mu ijk}=\Gamma_{\bar\mu\bar i\bar j\bar k}P_2^+,
\label{eq:m2gen}
\end{equation}
where $P_2^+=\fft12(1+\Gamma^{(2)})$ with
$\Gamma^{(2)}=\fft1{3!}\epsilon_{\mu\nu\rho}\Gamma^{\bar\mu\bar\nu\bar\rho}$.
In this case, the generalized connection takes the form
\begin{equation}
\Omega_\mu=\Omega_\mu^{\nu i}K_{\nu i},\qquad
\Omega_i=\ft23\partial_i\ln H_2\Gamma^{(2)}+\ft12\Omega_i^{jk}\hat T_{jk},
\end{equation}
where
\begin{equation}
\Omega_\mu^{\nu i}=-\ft43 H_2^{-1/2}\partial_i \ln H_2\delta_\mu^\nu,\qquad
\Omega_i^{jk}=\ft43\delta_{i[j}\partial_{k]}\ln H_2.
\end{equation}

First order integrability yields the generalized curvature
\begin{eqnarray}
M_{\mu\nu}\equiv\ft14\Rm_{\mu\nu}&=&0,\nonumber\\
M_{\mu i}\equiv\ft14\Rm_{\mu i}&=&\ft1{18}H_2^{-1/2}\left[
6(\partial_i\partial_j\ln H_2+2\partial_i \ln H_2\partial_j \ln H_2)
-(\partial\ln H_2)^2\delta_{ij}\right]K_{\mu j},\nonumber\\
M_{ij}\equiv\ft14\Rm_{ij}&=&\left[
-\ft13(\partial_l\partial_{[i\vphantom{]}}\ln H_2
-\ft13\partial_l\ln H_2\partial_{[i}\ln H_2)\delta_{j]k}
-\ft1{18}(\partial\ln H_2)^2\delta_{[i}^k\delta_{j]}^l
\right]\hat T_{kl},\qquad
\end{eqnarray}
which is similar in structure to that of the M5-brane (\ref{eq:m51order}).
The `missing' generator $K_{\mu ijk}$ of (\ref{eq:m2gen}) is obtained
by the commutation of $\hat T_{ij}$ with $\hat K_{\mu k}$.  This arises
in second order integrability via either $\D_i\Rm_{\mu j}$ or
$\D_\mu\Rm_{ij}$.  The general structure of the second order integrability
expressions are as follows:
\begin{eqnarray}
&&\kern-2em
M_{\mu\nu\lambda}=M_{\mu\nu\lambda}^{\rho i}K_{\rho i},\qquad
M_{\mu\nu i}=\ft12M_{\mu\nu i}^{jk}\hat T_{jk}+M_{\mu\nu i}^{\nu k}K_{\nu k},
\qquad
M_{\mu ij}=M_{\mu ij}^{\nu k}K_{\nu k}+\ft16M_{\mu ij}^{\nu klm}K_{\nu klm},
\nonumber\\
&&\kern-2em
M_{i\mu\nu}=0,\qquad
M_{i \mu j}=M_{i\mu j}^{\nu k}K_{\nu k}+\ft16M_{i\mu j}^{\nu klm}K_{\nu klm}
+\ft12M_{i\mu j}^{kl}\hat T_{kl},\qquad
M_{ijk}=\ft12M_{ijk}^{lm}\hat T_{lm}.
\end{eqnarray}
In this case, working to second order in integrability is sufficient to
guarantee closure of the holonomy algebra.  The group generated by
(\ref{eq:m2gen}) was identified in \cite{BDLW} to be
\begin{equation}
\Hol_{M2}=SO(8)_+ \ltimes 12\R^{2(8_s)}.
\end{equation}

It ought to be noted that the generalized connection $\Omega_M$
contains complete information about the generalized holonomy of
the spacetime, as the complete set of integrability conditions
(\ref{eq:nhint}) may be obtained through commutators and derivatives
of $\Omega_M$.  In this sense, the algebra of the holonomy group
can never be larger than the algebra obtained through the generators
in $\Omega_M$ itself.  However it can certainly be smaller.  This is
apparent for the M5-brane, where the $\Gamma^{(5)}$ generator is
absent in the generalized curvature $\Rm_{MN}(\Omega)$ and its
derivatives and also for the M2-brane, where $\Gamma^{(2)}$ is absent.
For these examples, and in fact for all vacua considered in
\cite{BDLW,BW}, the generators appearing in $\Omega_M$ and those
appearing in $\Rm_{MN}(\Omega)$ are nearly identical.  As a result,
the generalized holonomy group may be correctly identified at first
order in integrability, and the higher order conditions only serve to
complete the set of generators needed for closure of the algebra.

A different situation may arise, however, if for some reason (such
as accidental symmetries) a greatly reduced set of generators appear
in $\Rm_{MN}(\Omega)$.  In such cases, examination of first order
integrability may result in the misidentification of the actual
generalized holonomy group.  What happens here is that the algebra
of the curvature $\Rm_{MN}(\Omega)$ at a single point $p$ forms a
subalgebra of the holonomy algebra.  It is then necessary
to explore the curvature at all points $q$ connected by parallel
transport to $p$ in order to determine the actual holonomy algebra
itself.  Although this never occurs for Riemannian connections, we
demonstrate below that this incompleteness of first order integrability
does arise in the case of generalized holonomy.

\section{Higher order integrability and the squashed $S^7$}

For an example of the need to resort to higher order integrability
to characterize the generalized holonomy group $\Hol(\D)$, we turn to
Freund-Rubin compactifications of eleven-dimensional supergravity.
With vanishing gravitino, the Freund-Rubin ansatz for the $4$-form
field strength \cite{FR}
\begin{equation}
\label{Fleft}
F_{\mu \nu \rho \sigma} = 3m \epsilon_{\mu \nu \rho \sigma} , \quad
\mu=0,1,2,3,
\end{equation}
with $m$ constant and all other components vanishing, leads to
spontaneous compactifications of the product form AdS$_4 \times X^7$.  Here
$X^7$ is a compact, Einstein, Euclidean 7-manifold.  Decomposing the
eleven-dimensional Dirac matrices $\Gamma_M$ as
\begin{equation}
\Gamma_M = (\gamma_\mu \otimes 1 , \gamma_5 \otimes \Gamma_m),\qquad
\mu = 0,1,2,3,\qquad m=1,\ldots, 7,
\end{equation}
where $\gamma_\mu$ and $\Gamma_m$ are four- and seven-dimensional Dirac
matrices, respectively, and assuming the usual direct-product split
$\epsilon(x^\mu) \otimes \eta(y^m)$ for eleven-dimensional spinors, the
Killing spinor equation (\ref{eq:kse}) splits as
\begin{eqnarray}
\label{kads}
\D_\mu \epsilon &=& \left( \partial_\mu \
+\ft14 \omega_\mu{}^{\alpha \beta} \gamma_{\alpha \beta}
+ m \gamma_\mu \gamma_5 \right) \epsilon =0 ,  \\
\label{kleft}
\D_m \eta &=& \left( \partial_m \
+\ft14 \omega_m{}^{ab} \Gamma_{ab} - \ft{i}2 m \Gamma_m
\right) \eta =0 .
\end{eqnarray}
Since AdS$_4$ admits the maximum number of Killing spinors (four in this
case), the number $N$ of supersymmetries preserved in the compactification
coincides with the number of Killing spinors of the internal manifold $X^7$,
that is, with the number of solutions to the Killing spinor
equation (\ref{kleft}).  Therefore we only need to concern ourselves with
the Killing spinors on $X^7$.

An orientation reversal of $X^7$ or, alternatively, a sign reversal
of $F_{(4)}$, provides another solution to the equations of motion
(\ref{Einstein}), (\ref{4form}) and, hence, another acceptable
Freund-Rubin vacuum \cite{DNP83,DNP}.  For definiteness, we shall call
{\it left}-orientation the solution corresponding to the choice of sign of
$F_{(4)}$ in (\ref{Fleft}), that leads to the Killing spinor equation
(\ref{kleft}), and {\it right}-orientation the solution corresponding to the
opposite choice of sign of $F_{(4)}$:
\begin{equation} \label{Fright}
(\hbox{right})\qquad
F_{\mu \nu \rho \sigma} = -3m \epsilon_{\mu \nu \rho \sigma}, \quad
\mu=0,1,2,3,
\end{equation}
leading to the Killing spinor equation
\begin{equation} \label{kright}
(\hbox{right})\qquad
\D_m \eta = \left( \partial_m \
+\ft14 \omega_m{}^{ab} \Gamma_{ab} + \ft{i}2 m \Gamma_m \right) \eta =0 .
\end{equation}

{}From either (\ref{kleft}) or (\ref{kright}), we see that the generalized
connection $\D_m$ takes values in the algebra spanned by $\{ \Gamma_{ab},
\Gamma_a \}$ and therefore the generalized structure group is $\SO(8)$.
Notice, however, that both Killing spinor equations (\ref{kleft}) and
(\ref{kright}) share the same first order integrability condition
\cite{ADP,DNP}
\begin{equation} \label{int1}
M_{mn}\eta \equiv [\D_m,\D_n] \eta=\ft14\Rm_{mn}\eta
\equiv \ft14 \Cm_{mn}\eta=\ft14 C_{mn}{}^{ab} \Gamma_{ab} \eta= 0,
\end{equation}
where $ C_{mn}{}^{ab}$ is the Weyl tensor of $X^7$ (thus demonstrating
that, in this case the generalized curvature tensor is simply the Weyl
tensor).  Thus first order integrability is unable to distinguish between
left and right orientations on the sphere.  Then it might be possible
that spinors $\eta$ solving the integrability condition (\ref{int1}) will
only satisfy the Killing spinor equation for one orientation, that is,
satisfy (\ref{kleft}) but not (\ref{kright}) (or the other way around).
In fact, the skew-whiffing theorem \cite{DNP83,DNP} for Freund-Rubin
compactifications  proves that this will, in general, be the
case: it states that at most one orientation can give $N > 0$, with
the exception of the round $S^7$, for which both orientations give
maximal supersymmetry, $N=8$. Since the preserved supersymmetry $N$ is
given by the number of singlets in  the decomposition of the $\mathbf 8_s$
of $\SO(8)$ (the generalized structure group) under the generalized holonomy
group $\Hol(\D)$, it is then clear that, in general, each orientation must
have either a different generalized holonomy, or the same generalized
holonomy but a different decomposition of the $\mathbf 8_s$.

To illustrate this feature, consider compactifications on the squashed
$S^7$ \cite{DNP83,ADP}. This choice for $X^7$ has the topology of the
sphere, but the metric is distorted away from that of the round $S^7$;
it is instead the coset space $\SO(5) \times \SU(2) / \SU(2) \times
\SU(2)$ endowed with its Einstein metric \cite{DNP83,ADP}.  The
compactification on the left-squashed $S^7$ preserves $N=1$
supersymmetry whereas that on the right-squashed $S^7$ has $N=0$; put
another way, the integrability condition (\ref{int1}) has one
non-trivial solution, corresponding in turn to a solution to the Killing
spinor equation (\ref{kleft}) (making the left-squashed $S^7$ preserve
$N=1$), but not to a solution to (\ref{kright}), which in fact has no
solutions (yielding $N=0$ for the right-squashed $S^7$).  On the other
hand, an analysis of the Weyl tensor of the squashed $S^7$ shows that
there are only 14 linear combinations $\Cm_{mn}$ of gamma
matrices in (\ref{int1}), corresponding to the generators of $\mathrm
G_2$ \cite{ADP,DNP}.  Though appealing, $\mathrm G_2$ cannot be,
however, the generalized holonomy since the $\mathbf 8_s$ of $\SO(8)$
would decompose as $\mathbf 8_s \rightarrow \mathbf8\rightarrow
\mathbf 7+\mathbf 1$ under $\SO(8)\supset\SO(7)\supset\mathrm G_2$
regardless
of the orientation, giving $N=1$ for both left- and right-squashed solutions.
We thus conclude that in this case  the first order integrability condition
(\ref{int1}) is insufficient to determine the generalized holonomy.

The resolution to this puzzle is naturally given by higher order
integrability.  In the case of the squashed $S^7$, it turns out that
the second order integrability condition (\ref{eq:2int}) is
sufficient.  For a general Freund-Rubin internal space $X^7$ this
condition reads \cite{vNW}
\begin{equation} \label{int2}
M_{lmn}\eta\equiv\ft14 [\D_l,\Cm_{mn}]\eta=
\ft14\left(\nabla_lC_{mn}{}^{ab}\Gamma_{ab}\mp2imC_{mnl}{}^{a}\Gamma_a\right)
\eta = 0 \; ,
\end{equation}
the $-$ sign corresponding to the left solution, and the $+$ to
the right.  For the squashed $S^7$, we find that only 21 of the $M_{lmn}$
are linearly independent combinations of the Dirac matrices.  The
details are provided in the Appendix.  Following the notation of
\cite{ADP,DNP}, we split the index $m$ as $m=(0,i,\hat{i})$, with $i=1,2,3$,
$\hat{i}=4,5,6=\hat{1},\hat{2},\hat{3}$; then, with a suitable
normalization,  the linearly independent generators may be chosen to be
\begin{eqnarray} \label{G2}
&&\kern-2em
\Cm_{0i}=\Gamma_{0i}+\ft12\epsilon_{ikl} \Gamma^{\hat{k}\hat{l}},\quad
\Cm_{ij}= \Gamma_{ij}+\Gamma_{\hat{i}\hat{j}},\quad
\Cm_{i\hat{j}}=-\Gamma_{i\hat{j}}-\ft12\Gamma_{j\hat{i}}
+\ft12\delta_{ij}\delta^{kl}\Gamma_{k\hat{l}}
-\ft12\epsilon_{ijk}\Gamma^{0\hat{k}},\qquad\\
\label{SO7}
&&\kern-2em
M_{ij}=\Gamma_{\hat{i}\hat{j}} \mp \ft23\sqrt{5} im \epsilon_{ijk}
\Gamma^{\hat{k}},\quad
M_i= \Gamma_{0\hat{i}} \mp \ft23\sqrt{5} im \Gamma_i ,\quad
M= \delta^{kl}\Gamma_{k\hat{l}} \pm 2\sqrt{5}im \Gamma_0 ,
\end{eqnarray}
the $-$ sign in front of $m$ corresponding to the left solution and the
$+$ to the right.
Notice that there are 8 linearly independent generators in $\Cm_{i\hat j}$
of (\ref{G2}),
since $\delta^{kl} \Cm_{k \hat{l}} \equiv \Cm_{1\hat{1}} +\Cm_{2\hat{2}} +
\Cm_{3\hat{3}} =0$. The 14 generators $\Cm_{0i}$, $\Cm_{ij}$, $\Cm_{i\hat{j}}$
span $G_2$ \cite{ADP,DNP}, and are the same as those obtained from the first
integrability condition (\ref{int1}), while the 7 additional generators
$M_{ij}$, $M_i$, $M$ were not contained in (\ref{int1}). Taken together,
they generate the 21 dimensional algebra of $\SO(7)$, regardless of the
orientation, provided
\begin{equation}
m^2=\frac{9}{20} ,
\end{equation}
in agreement with the Einstein equation for the squashed $S^7$
\cite{DNP}.

The embedding of $\SO(7)$ into $\SO(8)$ is, however, different for
each orientation. We use $\SO(7)_-$ to denote the embedding corresponding
to the left solution and $\SO(7)_+$ the right.  While the spinor $\eta$
transforms as an $\mathbf 8_s$ of the generalized structure group $\SO(8)$,
the decomposition of the $\mathbf 8_s$ is different under left- and
right-squashing.  With our Dirac conventions, it turns out that
$\mathbf8_s \rightarrow \mathbf7+\mathbf1$ under $\SO(8)\supset\SO(7)_-$,
giving $N=1$ for the left-squashed $S_7$, while $\mathbf8_s \rightarrow
\mathbf8$ under $\SO(8)\supset\SO(7)_+$, giving $N=0$ for the
right-squashed $S^7$.

Since $\SO(7)$ is the subgroup of $\SO(8)$ that yields the correct
branching rules of the $\mathbf8_s$ of $\SO(8)$, we conclude that
second order integrability is sufficient in this case to identify all
generators of the Lie algebra of $\Hol(\D_m)$.  Hence the generalized
holonomy group of the Freund-Rubin compactification on the squashed
$S^7$ is precisely $\SO(7)$.  In this case, it is the embedding of
$\SO(7)$ in $\SO(8)$ (with corresponding spinor decomposition
$\mathbf8_s\to\mathbf7+\mathbf1$ or $\mathbf8_s\to\mathbf8$) that
determines the number of preserved supersymmetries.  This indicates that,
for generalized holonomy, knowledge of the holonomy group {\it and} the
embedding are both necessary in order to understand the number of
preserved supersymmetries.  While this was already observed in
\cite{BDLW,Hull:2003mf} for non-compact groups, here we see that this
is also true when the generalized holonomy group is compact.

The analysis of the squashed $S^7$, along with the brane solutions
of the previous section, highlights several features of generalized
holonomy.  For the squashed $S^7$, the generalized holonomy algebra
is in fact larger than that generated locally by the Weyl curvature
at a point $p$.  In this case, the algebra arising from lowest order
integrability is already closed, but is only a subalgebra of the
correct holonomy algebra.  It is then mandatory to examine the second
order integrability expression (\ref{int2}) in order to identify the
generalized holonomy group.  On the other hand, for the M2- and
M5-branes, lowest order integrability, while lacking a complete set
of generators, nevertheless closes on the correct holonomy algebra, and
no really new information is gained at higher order.

Of course, in all cases, complete information is contained in the
generalized connection $\Omega_M$ itself.  However, examination of
$\Omega_M$ directly can be misleading, as it may contain gauge
degrees of freedom, which are unphysical.  This is most clearly
seen in the case of the round $S^7$, where $\Omega_m=\omega_m^{ab}
\Gamma_{ab}-2im\Gamma_m$ is certainly non-vanishing, while the
generalized curvature $\Rm_{mn}$, given by the Weyl tensor, is
completely trivial.

For generalized holonomy to be truly useful, we believe it ought
to go beyond simply a classification scheme, and must yield methods
for constructing new supersymmetric backgrounds.  In much the same
way that the rich structure of Riemannian holonomy teaches us a
great deal about the geometry of Killing spinors on Riemannian
manifolds, we anticipate that the formal analysis of generalized
holonomy via connections on Clifford bundles may one day lead to
a similar expansion of knowledge of supergravity structures and
manifolds with fluxes.  While much remains to be done, as we have
only highlighted a few examples, we hope that a more complete
understanding of higher order integrability for the curvature of
generalized connections will soon lead to a better appreciation
of the geometry behind generalized holonomy.

\section*{Acknowledgments}
We are grateful for discussions with M.~Duff, who provided the initial
suggestion that the generalized holonomy of the squashed $S^7$ must
be $\SO(7)_\pm$ with corresponding decomposition of the $\mathbf8_s$ of
$\SO(8)$ \cite{Duff:2002rw}.  We also wish to thank I.~Bandos,
J.~A.~de~Azc\'arraga and P.~de~Medeiros for useful conversations.
This work was supported in part by the US Department of Energy under
grant DE-FG02-95ER40899, the Spanish Ministerio de Educaci\'on y
Ciencia and EU FEDER funds under grant BFM2002-03681, the Generalitat
Valenciana (Grupos
03/124) and the EU network MRTN-CT-2004-005104 `Forces
Universe'. O.V.~wishes to thank the Generalitat Valenciana for his FPI
research grant and for funding of his stay at the MCTP, and to
M.~Duff and the MCTP for kind hospitality.

\appendix

\section{Second order integrability for the squashed $S^7$}

In this Appendix we present the details of the derivation of the
linearly independent generators (\ref{G2}) and (\ref{SO7}) of the
generalized holonomy group $\Hol(\D_m)=SO(7)$ of the squashed $S^7$,
associated to the second order integrability condition (\ref{int2}).
For convenience, we rewrite (\ref{int2}) with a modified normalization
\begin{equation} \label{second}
M_{abc} = 5\left( \sqrt{5}
     \nabla_a C_{bcde} \Gamma^{de} \
     -m' C_{bcad} \Gamma^d \right)    \; ,
\end{equation}
where we have defined
\begin{equation}
m'=2\sqrt{5} i m ,
\end{equation}
and have chosen the $-$ sign in front of $m'$ for definiteness.

To obtain $M_{abc}$, we have computed both the Weyl tensor $C_{bcad}$
(given in \cite{DNP}) and its covariant derivative $\nabla_aC_{bcde}$.
We obtain, for the non-vanishing generators:
\begin{eqnarray}
\label{1}
M_{00j} &=& 4 \Gamma_{0 \hat{j}} - \epsilon_{jkl}
\Gamma^{k \hat{l}} -2m' \Gamma_j \; , \\
\label{2}
M_{00\hat{j}} &=& 4 \Gamma_{0j} +
 \epsilon_{jkl}  \Gamma^{kl} +2m'\Gamma_{\hat{j}} \; , \\
\label{3}
M_{0ij} &=& 2 \epsilon_{ijk} \Gamma^{0 \hat{k}} +
\Gamma_{i \hat{j}}-\Gamma_{j \hat{i}} \; , \\
\label{4}
M_{0i\hat{j}} &=& -\epsilon_{ijk} \Gamma^{0k} +
\Gamma_{ij}-3\Gamma_{\hat{i} \hat{j}} + m' \epsilon_{ijk}
\Gamma^{\hat{k}} \; , \\
\label{5}
M_{0\hat{i}\hat{j}} &=& - 3\Gamma_{i \hat{j}}
+3\Gamma_{j\hat{i}} -2m' \epsilon_{ijk} \Gamma^{k} \; , \\
\label{6}
M_{h0j} &=& \epsilon_{hjk} \Gamma^{0 \hat{k}}
+2 \Gamma_{h\hat{j}} + \delta_{hj} \delta^{kl} \Gamma_{k\hat{l}}
+\Gamma_{j\hat{h}} +2m' \delta_{hj} \Gamma_{0} \; ,  \\
\label{7}
M_{h0\hat{j}} &=& -\epsilon_{hjk} \Gamma^{0k}
+ \Gamma_{hj} + 3 \Gamma_{\hat{h}\hat{j}}
-m' \epsilon_{hjk} \Gamma^{\hat{k}} \; , \\
\label{8}
M_{hij} &=&  \delta_{hi} \Gamma_{0\hat{j}}
-\delta_{hj} \Gamma_{0\hat{i}} + 4\epsilon_{ij}{}^k \Gamma_{h\hat{k}}
-\epsilon_{hij} \delta^{kl} \Gamma_{k\hat{l}}
- \epsilon_{ij}{}^k \Gamma_{k\hat{h}}
+ 2m'(\delta_{hj} \Gamma_i - \delta_{hi} \Gamma_j) \; , \\
\label{9}
M_{hi\hat{j}} &=& 2\delta_{hi} \Gamma_{0j}
+\delta_{ij} \Gamma_{0h} +\delta_{hj} \Gamma_{0i}
+ (2 \epsilon_{jkl} \delta_{hi} -\ft12 \epsilon_{hkl}
\delta_{ij}  -\ft12 \epsilon_{ikl} \delta_{hj}) \Gamma^{kl}
-3\epsilon_{hi}{}^k \Gamma_{\hat{k} \hat{j}}\qquad \nonumber \\
&&+ m'(2\delta_{hi} \Gamma_{\hat{j}} -
\delta_{ij} \Gamma_{\hat{h}}
+\delta_{hj} \Gamma_{\hat{i}}) \; ,  \\
\label{10}
 M_{h\hat{i}\hat{j}} &=&  3\delta_{hi} \Gamma_{0\hat{j}}
-3\delta_{hj} \Gamma_{0\hat{i}} + 3 \epsilon_{hi}{}^k \Gamma_{k\hat{j}}
-3\epsilon_{hj}{}^k \Gamma_{k\hat{i}}
+ 2m'(\epsilon_{hij} \Gamma_0 + \delta_{hj} \Gamma_i
-\delta_{hi} \Gamma_j ) \; , \\
\label{11}
M_{\hat{h}0j} &=& -6 \Gamma_{\hat{h}\hat{j}}
+ 2m' \epsilon_{hjk} \Gamma^{\hat{k}} \; , \\
\label{12}
M_{\hat{h}0\hat{j}} &=& 3\Gamma_{j \hat{h}}
-3 \delta_{hj} \delta^{kl}\Gamma_{k\hat{l}}
-m' (2\delta_{hj} \Gamma_0 + \epsilon_{hjk} \Gamma^k) \; , \\
\label{13}
M_{\hat{h}ij} &=& 6 \epsilon_{ij}{}^k \Gamma_{\hat{k}\hat{h}} +2m'( \delta_{hj}
\Gamma_{\hat{i}}-\delta_{hi} \Gamma_{\hat{j}}) \; , \\
\label{14}
M_{\hat{h}i\hat{j}} &=& 3\delta_{hj} \Gamma_{0 \hat{i}}
-3\delta_{ij} \Gamma_{0 \hat{h}}-
3\delta_{hi}\epsilon_{jkl}\Gamma^{k\hat{l}} -
3\epsilon_{ij}{}^l \Gamma_{h\hat{l}}\nonumber\\
&&+m'( -\epsilon_{hij} \Gamma_0 -2\delta_{hj} \Gamma_i
+\delta_{ij} \Gamma_{h} -\delta_{hi} \Gamma_j) \; , \\
\label{15}
M_{\hat{h}\hat{i}\hat{j}} &=&  6\delta_{hj} \Gamma_{0i}
 -6\delta_{hi} \Gamma_{0j}
-6 \epsilon_{ij}{}^k\Gamma_{kh}
+4m'(\delta_{hj} \Gamma_{\hat{i}}
-\delta_{hi} \Gamma_{\hat{j}}) \; .
\end{eqnarray}
Not all the generators included in (\ref{1})--(\ref{15}) are linearly
independent, however. After all, they are built up from Dirac
matrices $\{ \Gamma_{ab}, \Gamma_a \}$, that is, from generators of
$\SO(8)$, so at most 28 can be linearly independent.

In fact, only 21 linearly independent generators are
contained in (\ref{1})--(\ref{15}), as we will now show.
Some redundant generators are
straightforward to detect, since the Bianchi identities for the Weyl
tensor, $\nabla_{[a} C_{bc]de} =0$ and $C_{[bca]d}=0$ place the restrictions
\begin{equation}
M_{[abc]}=0 \; .
\end{equation}
Further manipulations show that only the
generators (\ref{9}) and (\ref{14}) are relevant, the rest being linear
combinations of them. The generators  (\ref{2}), (\ref{4}), (\ref{7}),
(\ref{11}), (\ref{13}) and (\ref{15}) are obtained from (\ref{9}):
\begin{eqnarray}
M_{00\hat{j}} &=& \ft15 \delta^{kl}
( M_{kl\hat{j}} + M_{kj\hat{l}} + M_{jk\hat{l}} ) \; , \\
M_{0i\hat{j}} &=& \ft15 \epsilon_{[i \mid}{}^{kl}
(4 M_{k \mid j ] \hat{l}} -M_{\mid j ] k\hat{l}})
-\ft15 \epsilon_{ij}{}^k \delta^{lm} M_{lm\hat{k}} \; , \\
M_{i0\hat{j}} &=& -\ft15 \epsilon_{[i \mid}{}^{kl}
(M_{k \mid j ] \hat{l}} -4M_{\mid j ] k\hat{l}})
-\ft15 \epsilon_{ij}{}^k \delta^{lm} M_{lm\hat{k}} \; , \\
M_{\hat{j}0i} &=& -\epsilon_{[i \mid}{}^{kl}
(M_{k \mid j ] \hat{l}} -M_{\mid j ] k\hat{l}}) \; , \\
M_{\hat{h}ij} &=& M_{ji\hat{h}} - M_{ij\hat{h}} \; , \\
M_{\hat{h}\hat{i}\hat{j}} &=& \ft15 ( M_{hi\hat{j}} -M_{hj\hat{i}}
+ M_{ih\hat{j}} - M_{jh\hat{i}} ) -\ft45 \delta^{kl}
(\delta_{hi} M_{kl\hat{j}} - \delta_{hj} M_{kl\hat{i}} ) \; ,
\end{eqnarray}
while  (\ref{1}), (\ref{3}), (\ref{5}),
(\ref{6}), (\ref{8}), (\ref{10})
and (\ref{12}) are linear combinations of (\ref{14}):
\begin{eqnarray}
M_{00j} &=& \ft13 \delta^{kl}
(M_{\hat{k}j\hat{l}}-M_{\hat{j}k\hat{l}} ) \; , \\
M_{0hj} &=& -\ft13 \epsilon_h{}^{kl} (M_{\hat{k}l\hat{j}} +3
M_{\hat{j}k\hat{l}} ) +\ft13 \epsilon_j{}^{kl}
 \left(M_{\hat{k}l\hat{h}} +3M_{\hat{h}k\hat{l}} \right) \; , \\
M_{0\hat{h}\hat{j}} &=& \epsilon_h{}^{kl} (M_{\hat{k}l\hat{j}} +2
M_{\hat{j}k\hat{l}} ) - \epsilon_j{}^{kl}
 (M_{\hat{k}l\hat{h}} +2M_{\hat{h}k\hat{l}} ) \; , \\
M_{h0j} &=& -\ft16 \epsilon_h{}^{kl} (2 M_{\hat{k}l\hat{j}} +5
M_{\hat{j}k\hat{l}} ) +\ft16 \epsilon_j{}^{kl}
M_{\hat{h}k\hat{l}} \; , \\
M_{hij} &=& \ft12 \delta^{kl} \left( \delta_{hi}
( M_{\hat{k}l\hat{j}} - 2M_{\hat{j}k\hat{l}} ) -\delta_{hj}
( M_{\hat{k}l\hat{i}} - 2M_{\hat{i}k\hat{l}} ) \right) \nonumber \\
&&+ \ft73 ( M_{\hat{h}i\hat{j}} - M_{\hat{h}j\hat{i}} ) +
M_{\hat{i}j\hat{h}} - M_{\hat{j}i\hat{h}} - \ft23 \epsilon_h{}^{kl}
\epsilon_{ij}{}^m ( M_{\hat{k}l\hat{m}}+4M_{\hat{m}k\hat{l}} )\;,\\
M_{h\hat{i}\hat{j}} &=&  M_{\hat{i}h\hat{j}} -  M_{\hat{j}h\hat{i}} \; , \\
M_{\hat{h}0\hat{j}}&=&\epsilon_h{}^{kl} (M_{\hat{k}l\hat{j}} +
M_{\hat{j}k\hat{l}} ) -\epsilon_j{}^{kl}
M_{\hat{h}k\hat{l}} \; ,
\end{eqnarray}

Moreover, both (\ref{9}) and (\ref{14}) contain redundant generators. The
following combinations obtained from (\ref{9}):
\begin{eqnarray}
\Cm_{0i} &=& \ft16 \delta^{kl} M_{ik\hat{l}} \; , \\
\Cm_{ij} &=& -\ft1{30} \epsilon_{[i \mid}{}^{kl}
(M_{k \mid j ] \hat{l}} -9M_{\mid j ] k\hat{l}})
-\ft1{30} \epsilon_{ij}{}^k \delta^{lm} M_{lm\hat{k}} \; , \\
M_{ij} &=& \ft16 M_{\hat{j}0i}= -\ft16 \epsilon_{[i \mid}{}^{kl}
(M_{k \mid j ] \hat{l}} -M_{\mid j ] k\hat{l}}) \;
\end{eqnarray}
(the expressions of which in terms of Dirac matrices are the first two
equations in (\ref{G2}) and the first equation in (\ref{SO7}), respectively)
are linearly independent.  Thus (\ref{9}) [and so (\ref{2}),
(\ref{4}), (\ref{7}), (\ref{11}), (\ref{13}) and (\ref{15})]
can be uniquely written in terms of them:
\begin{eqnarray}
M_{hi\hat{j}} &=& 2\delta_{hi} \Cm_{0j}+ \delta_{ij} \Cm_{0h} + \delta_{hj}
\Cm_{0i}
+ (2 \epsilon_j{}^{kl} \delta_{hi} -\ft12 \epsilon_h{}^{kl}
\delta_{ij}  -\ft12 \epsilon_i{}^{kl} \delta_{hj}) \Cm_{kl}
\nonumber \\
&&- 3\delta_{hi} \epsilon_j{}^{kl} M_{kl} -3 \epsilon_{hi}{}^k M_{kj} \; .
\end{eqnarray}

Similarly, the following combinations contained in (\ref{14}):
\begin{eqnarray}
\Cm_{i \hat{j}} &=& \ft13 \epsilon_i{}^{kl} M_{\hat{j}k\hat{l}} -
\ft16  \epsilon_j{}^{kl} (M_{\hat{k}l\hat{i}} +
M_{\hat{i}k\hat{l}} ) \; , \\
M_i &=& \ft1{12} \delta^{kl} (M_{\hat{k}l\hat{i}} - 2
M_{\hat{i}k\hat{l}}) \; , \\
M &=& -\ft16 \epsilon^{hij} M_{\hat{h}i\hat{j}} \;
\end{eqnarray}
(which can be written in terms of Dirac matrices as in the final equation
of (\ref{G2}) and the last two equations of (\ref{SO7}), respectively)
are linearly independent.  Hence (\ref{14}) [and so (\ref{1}), (\ref{3}),
(\ref{5}), (\ref{6}), (\ref{8}), (\ref{10}) and (\ref{12})] can be uniquely
written in terms of them:
\begin{eqnarray}
M_{\hat{h}i\hat{j}} &=& 6 \delta_{hi} \epsilon_j{}^{kl} \Cm_{k\hat{l}}
-2\epsilon_{ij}{}^k( \Cm_{k\hat{h}}-2 \Cm_{h\hat{k}}) \nonumber \\
&&+ 6 \delta_{hj} M_i  + 3\delta_{ih} M_j - 3\delta_{ij} M_h
-\epsilon_{hij} M \; .
\end{eqnarray}

In summary, the linearly independent generators associated to the second
order integrability condition (\ref{int2}) are the 21 linearly
independent generators (\ref{G2}) and (\ref{SO7}), namely
$\{ \Cm_{0i}$, $\Cm_{ij}$, $\Cm_{i\hat{j}}$, $M_{ij}$, $M_i, M \}$
(notice that $\Cm_{i\hat{j}}$ contains 8 generators, since it is traceless),
which close into an algebra whenever $m^2$ takes the value required by the
equations of motion,  $m^2 = \frac{9}{20}$. Since the only condition for
the generators to close the algebra is placed on $m^2$, they
will close regardless of the orientation ({\it i.e.}, of the
sign of $m$). In fact, they generate the 21-dimensional algebra of
$SO(7)$, for both orientations.

Note that, by further choosing linear combinations of (\ref{G2}),
the 14 generators $\{\Cm_{0i}$, $\Cm_{ij}$, $\Cm_{i\hat j}\}$ of
$\mathrm G_2$ may be re-expressed in symmetric form
\begin{eqnarray}
&&\Gamma_{1\hat1}-\Gamma_{2\hat2},\qquad\Gamma_{1\hat1}-\Gamma_{3\hat3},
\nonumber\\
&&\Gamma_{0\hat i}+\Gamma_{j\hat k},\qquad
\Gamma_{0\hat i}+\Gamma_{\hat jk},\qquad(i,j,k=123,231,312)
\nonumber\\
&&\Gamma_{0i}+\Gamma_{\hat j\hat k},\qquad
\Gamma_{0i}-\Gamma_{jk},\qquad(i,j,k=123,231,312).
\label{eq:g2can}
\end{eqnarray}
The 7 additional generators $\{M_{ij}$, $M_i$, $M\}$ of (\ref{SO7})
extending (\ref{eq:g2can}) to $\SO(7)$ may also be simplified in
appropriate linear combinations.  One possible set of generators is
given by:
\begin{eqnarray}
&&\Gamma_{1\hat1}\pm i\Gamma_0,\nonumber\\
&&\Gamma_{0\hat i}\mp i\Gamma_{i},\nonumber\\
&&\Gamma_{\hat j\hat k}\mp i\Gamma_{\hat i},\qquad(i,j,k=123,231,312).
\end{eqnarray}
%

\bibliographystyle{amsplain}

\end{document}